# A nonlinear metasurface combining telecom-range intersubband transitions in GaN/AlN quantum wells with resonant plasmonic antenna arrays

JAN MUNDRY,[1,*] FLORIAN SPREYER,[2] VALENTIN JMERIK,[3] SERGEY IVANOV,[3] THOMAS ZENTGRAF,[2] AND MARKUS BETZ[1]

[1]*Experimentelle Physik 2, TU Dortmund University, 44227 Dortmund, Germany*
[2]*Department of Physics, Paderborn University, 33098 Paderborn, Germany*
[3]*Ioffe Institute, St. Petersburg, Russia*

*\*jan.mundry@tu-dortmund.de*

**Abstract:** We realize and investigate a nonlinear metasurface taking advantage of intersubband transitions in ultranarrow GaN/AlN multi-quantum well heterostructures. Owing to huge band offsets, the structures offer resonant transitions in the telecom window around 1.55 µm. These heterostructures are functionalized with an array of plasmonic antennas featuring cross-polarized resonances at these near-infrared wavelengths and their second harmonic. This kind of nonlinear metasurface allows for substantial second-harmonic generation at normal incidence which is completely absent for an antenna array without the multi-quantum well structure underneath. While the second harmonic is originally radiated only into the plane of the quantum wells, a proper geometrical arrangement of the plasmonic elements permits to redirect the second-harmonic light to free-space radiation, which is emitted perpendicular to the surface.

## 1. Introduction

Frequency conversion processes such as second-harmonic generation (SHG) are integral to nonlinear optics, which is the bedrock of modern photonic science and technology. The wide field of application covers e.g. the generation of coherent ultraviolet light [1], supercontinuum light generation [2], nonlinear spectroscopy [3], the generation of entangled photon pairs [4] and many more. The nonlinear response of naturally occurring materials is inherently weak such that, e.g., phase-matching techniques [5,6] are required to realize significant nonlinear response. Therefore researchers are currently targeting the synthesis of novel materials that offer substantial nonlinear response even from thin layers [7-9]. Plasmonic metasurfaces offer a large potential for nonlinear optical interactions. They combine deep-subwavelength thickness with the possibility to engineer the shape of these structures to resonantly enhance the light-matter interaction at certain frequencies. It is strongly desirable to combine plasmonic structures with semiconductor-based materials to further enhance nonlinear susceptibilities [9,10]. Doped quantum well [11] and doped multi-quantum well (MQW) structures [12] are particularly suited to tailor the nonlinear optical response in the infrared. By proper choice of the material combinations and well widths, intersubband transitions (ISBTs) can be quantum-engineered to match certain optical frequencies. At the same time, MQW structures can feature a broken inversion symmetry required for SHG [13]. However, the excitation of ISBTs requires vertical (z-polarized) electric field components that are absent, e.g., for free-space illumination at normal incidence. Therefore the combination with plasmonic structures partially converting normally impinging light into such z-polarized components in the MQW volume is particularly beneficial. Lee et al. [9] have demonstrated record-high levels of SHG efficiencies at such ISBTs by functionalizing AlGaAs/GaAs heterostructures with antenna arrays featuring resonances for both fundamental ($\omega$) and second harmonic ($2\omega$) frequencies. Tuning this concept to the more practical near-infrared regime requires the use of compounds with larger band offsets and narrower quantum wells. Group III-nitrides are a good platform to meet these requirements as a GaN/AlN heterojunction offers a conduction band offset of ~2 eV [14]. Based on this material system in combination with resonant plasmonic resonators, Wolf et al. [15] have demonstrated enhanced SHG efficiencies for fundamental radiation around 3.5 µm. However, tuning to shorter wavelengths is hampered by the finite depth of the confining potential in the conduction band.

In this paper, we investigate a nonlinear metasurface operating with telecom range ISBTs in ultranarrow GaN/AlN MQWs and find substantial SHG for femtosecond pulses at 1.55 µm central wavelength at normal incidence. Hexagonal GaN-based heterostructures exhibit both a crystalline inversion asymmetry and a structural inversion asymmetry arising, e.g., from internal polarization fields and, therefore allow for SHG. While the first intersubband transition is centered at 1.55 µm in quantum wells with a thickness of 1.8 nm, any potential second harmonic is resonant only to bound-to-continuum transitions. The heterostructure is functionalized with a nominally inversion-symmetric antenna array. Specifically, we use a periodic array of rod antennas with their length and width chosen to support cross-polarized plasmonic resonances at the driving wavelength of 1.55 µm and its second harmonic. Resulting, we find second harmonic emission to occur in the plane of the MQWs guided through an AlN buffer layer underneath that acts as a multimode waveguide for the second harmonic. For an appropriate choice of the lateral antenna distances within the array, the second harmonic is coupled into free-space radiation and emitted perpendicular to the surface with a polarization rotated by 90 degrees with respect to the incident fundamental light.

## 2. Sample overview and optical properties

The heterostructure of this study, schematically depicted in Fig. 1(a), is grown by plasma-assisted molecular beam epitaxy (Compact21T, Riber) on a double-side polished c-sapphire wafer (c-Al$_2$O$_3$). Initially, a 1.6 µm thick AlN layer was grown using two-stage metal-modulated epitaxy to reduce the threading dislocation density to mid-$10^9$ cm$^{-2}$ and to achieve an atomically smooth surface, as detailed in [16]. Then, 25 GaN QWs with a thickness of 1.8 nm, separated by 5nm thick AlN barriers layers, were formed using the sub-monolayer digital alloying method described in [17]. While the GaN QWs are nominally undoped, the AlN barriers were grown with Si-doping with a concentration of n = 1.5 x $10^{19}$ cm$^{-3}$. The n-doping gives rise to a two-dimensional electron gas with an expected density of n = 4 x $10^{12}$ cm$^{-3}$, corresponding to a calculated Fermi energy of $E_F = (\pi\hbar^2/m^*) \cdot n$ = 90 meV above the minimum of the first subband (note that m* = 0.20 m$_0$ for wurtzite GaN [18]). The MQW structure was capped and protected by a 5 nm AlN layer. As a result, the structures offer a large density of states for resonant ISBTs centered at ~0.8 eV. Our previous study on a nominally almost identical, cubic, heterostructure show that QWs of the present width exhibit ISBTs around 0.8 eV [19]. This finding is consistent with bandstructure simulations performed with the commercial nextnano simulation package. The resonance is clearly broader than the spectrum of the nearly transform-limited ~50 fs laser pulses of our experiment.

Fig. 1(a) and 1(b) illustrate the plasmonic elements deposited on top of the GaN/AlN heterostructure. They are made of gold and fabricated by standard electron-beam lithography. After the lithography, the Polymethylmethacrylat (PMMA) photoresist is developed and 30 nm of gold is deposited by electron beam evaporation onto the sample. In the last step, the plasmonic structures are laid bare via a lift-off process where the PMMA mask is removed. We find antenna length of l = 300 nm up to l = 350 nm to best match the profile of the ISBT and the available femtosecond pulses from our optical parametric amplifier (OPA) source such that we restrict the present paper to such antenna lengths. We note that we have studied several other antenna lengths and found the nonlinear response to decrease as expected since their resonance shifts away from the laser spectrum and/or the ISBTs.

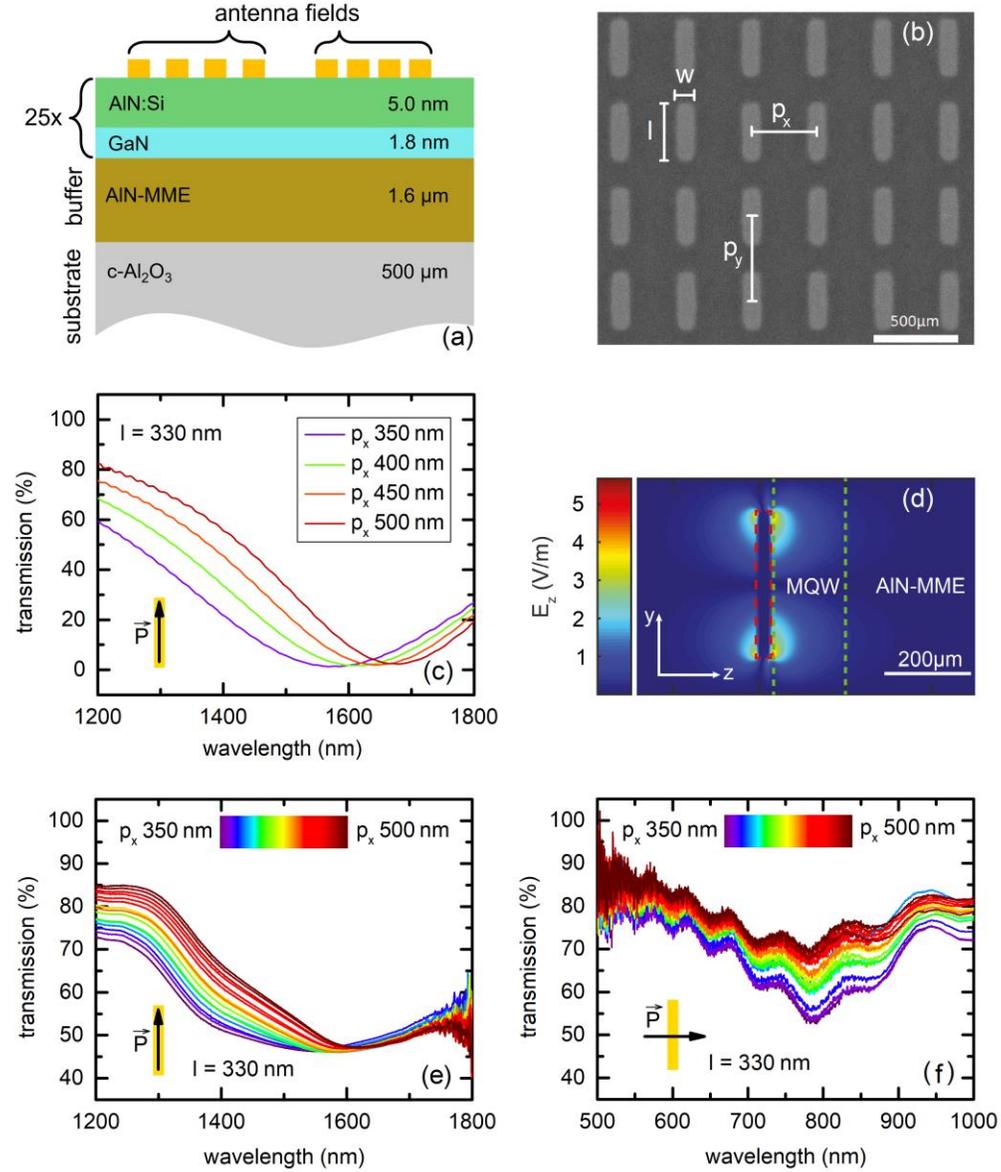

Fig. 1. (a) Schematic illustration of the semiconductor heterostructure: The AlN buffer layer and 25 GaN/AlN quantum wells are grown on a sapphire substrate. The antenna arrays are fabricated on top of the heterostructure. (b) Top view image of the plasmonic antennas taken on a scanning electron microscope. Width (w), length (l) and spacings in x- and y-direction ($p_x$ and $p_y$) are included. (c) Calculated transmission spectra for different periods $p_x$ of plasmonic antenna arrays on an AlN-substrate. (d) Calculated electric field close to the antenna. The antenna is indicated by a dashed red line and the MQW area is located between the two dashed green lines. (e) Transmission spectra of several arrays for y-polarized (longitudinal) incident light in the near-infrared. (f) Transmission spectra of the corresponding antenna arrays for x-polarized (transversal) incident light in the visible range.

Fig. 1(b) shows a top view image of the plasmonic antennas taken on a scanning electron microscope that contains the notation for the dimensions of the antennas. They have a height of h = 30 nm and a width of w = 80 nm. While the separation along the y-axis is set to a constant value of $p_y$ = 500 nm, we systematically vary the lateral distance $p_x$ between 350 and 500 nm. Fields of 100 µm x 100 µm are filled with a periodic array of antennas with fixed $p_x$. The panel (c) in Fig. 1 shows calculated transmission spectra for different periods $p_x$ of plasmonic antenna arrays on an AlN-substrate. The simulations are done by a plane wave excitation with y-polarized light performed with CST Studio Suite. Furthermore, the z-components of the calculated electric field close to the antenna are extracted and illustrated in panel (d). It is shown, that strong near-fields are coupled into the MQW region of the stacked AlN/GaN layers. The panels (e) and (f) of Fig. 1 display the transmission spectra of several arrays for y-polarized (longitudinal) light in the near-infrared as well as for x-polarized (transversal) light in the visible range. They are indicative of plasmonic resonances at 1.55 µm and its second harmonic at 775 nm wavelength. The measured resonances at the fundamental wavelength in panel (e) are in line with the corresponding design wavelengths shown in panel (c). Details of the spectra such as the widths of the resonances depend on the lateral distance $p_x$ which indicates significant coupling between adjacent antennas [20,21].

### 3. Experimental approach

The experimental setup is schematically shown in Fig. 2. It is designed for polarization-resolved transmission measurements. The main focus of the experiment is the detection and spectral analysis of the second and third harmonic signals generated in transmission geometry. A modelocked amplified laser system (Coherent RegA 9040) in combination with an OPA (Coherent OPA 9850) is used as a pump source. It delivers a train of linearly polarized, nearly transform-limited ~50 fs pulses at a repetition rate of 250 kHz. While this source is somewhat tunable, we choose a fixed center wavelength of 1550 nm such that the pulses have a good overlap with the ISBTs and the antennas' resonance. If not stated otherwise, the measurements are performed in a single pass geometry with perpendicular incidence and longitudinal light polarization with respect to the orientation of the antennas. The sample is clamped to a tiltable holder and illuminated from the backside. Using a half-wave-plate and a polarizing beam splitter cube, a portion of up to a few hundred microwatts is focused down to a beam waist of approximately 30 µm using a lens with a focal length of 50 mm. This value corresponds to the full with at half maximum and is measured with a commercial scanning-slit beam profiler (Thorlabs BP209-IR). The detection unit for the nonlinear signals consists of a sensitive camera (Atik 414EX, for imaging purposes) and a monochromator with a cooled CCD array (Andor Cornerstone, DU420-0E), for spectrally resolved measurements). The additional use of a suitable short-pass filter with high transmissivity at the second and third harmonic wavelength ensures the suppression of the fundamental pump beam. A linear polarizer and an achromatic half-wave-plate in front of the detection unit allow for polarization-resolved measurements.

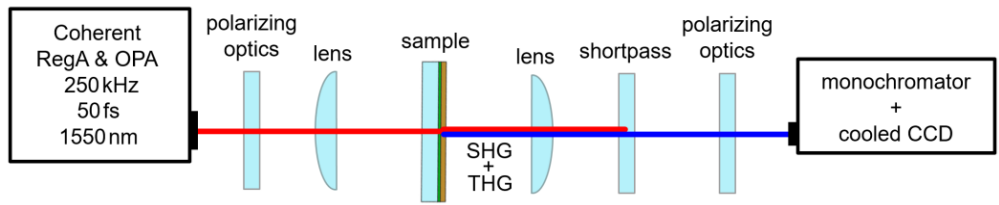

Fig. 2. Sketch of the experimental setup. The fundamental laser beam is focused down to the antenna arrays, passing the sample from the backside. An additional short-pass filter and polarization optics allow for spectrally and polarization-resolved measurements using a monochromator with a cooled CCD array.

## 4. Results and discussion

### 4.1 Pump power dependence

In the first step, we provide an example of the nonlinear optical response of the metasurface. Fig. 3(a) and (b) contain spectra recorded for 1550 nm fundamental pulses of average powers ranging from 50 µW to 400 µW. The upper power limit is chosen to be smaller than the threshold for permanent, visible damage to the plasmonic antennas occurring at power levels around 3 mW, referring to a peak power density of 420 W/cm². The length of the antennas is l = 330 nm and the separation in the x-direction is $p_x$ = 400 nm. Under these conditions, substantial SHG is observed, cf. panel (b). It is almost as strong as the THG signal, cf. panel (a). While THG is expected for the C2 symmetry of the antennas, the observation of substantial SHG cannot arise from the antenna array solely. The spectrum of the SHG shows an unexpected structure with two peaks. It arises from the modal structure of the AlN buffer layer and will be discussed later in the manuscript. Fig. 3(c) and (d) show the quantitative analysis of SHG and THG as a function of the fundamental pump power. The pump-power dependence of the THG (SHG) reveals a cubic (quadratic) behavior, as expected for these nonlinear processes. The deviation between the experimental data and the parabola fit for low SHG intensity in panel (d) may arise from residual 800 nm light from the regenerative amplifier system. Furthermore, in comparison to similar measurements on MQWs found in literature [9,22], no pronounced saturation of SHG intensity is visible in our study. We attribute this finding to the rather strong doping. For a Fermi energy of ~100 meV, depletion of the lowest subband is hard to achieve.

In the remainder of the manuscript, we want to provide comprehensive evidence that the unexpected SHG arises from an interplay between the ISBTs and the antenna array. As a first aspect of such evidence, we show that none of the two ingredients alone allows for SHG. To this end, we investigate a reference sample where no MQW is present, i.e., the antenna arrays are deposited onto the sapphire substrate. In this case, the SHG signal is below the detection limit of our setup. This holds true for all antenna arrays irrespective of the value of $p_x$. THG from the reference sample is about a factor of ~5 smaller compared to the results shown in Fig. 3. We also investigate SHG/THG on a bare GaN/AlN heterostructure without any plasmonic structures. Again, SHG is below the detection limit of our experiment. THG is reduced by two orders of magnitude when compared to the results in Fig. 3. Therefore we conclude that the SHG of this nonlinear metasurface is based on the interplay of ISBTs in GaN/AlN MQWs with matched plasmonic antenna arrays. The absence and/or mismatch of one component leads to the practically complete disappearance of the SHG signal.

One of the most interesting aspects is the true output power of second harmonic radiation from the nonlinear metasurface. The measurable value strongly depends on the choice of the plasmonic elements that permit conversion of the second harmonic light to free-space radiation. The emission is measured perpendicular to the surface in the far-field using the CCD from the spectrometer. For calibration purpose, a thin BBO crystal is used at the original position of the sample. This allows second-harmonic power measurements with ordinary commercial devices leading to an accurate power-per-count calibration. The most intense second harmonic output for a constant pump power of 400 µW is observed for antennas with l = 340 nm and $p_x$ = 420 nm and equals about 10 fW. The calculated power conversion efficiency $\eta_{SHG} = P_{SHG}/P_{Pump}$ is therefore $2.5 \cdot 10^{-11}$. It is well possible that the actual conversion efficiency is larger. The experiment is designed to detect the second harmonic generated at the excitation spot with a microscope objective with a numerical aperture of 0.25. Due to the spectrally wide pump beam (full width at half maximum is $\Delta\lambda \approx 80$ nm at 1550 nm central wavelength) the spectral components of the SHG, converted into transmission geometry from the antenna arrays, are emitted into slightly different directions with respect to the surface normal. With a numerical aperture of 0.25, most of them will still be collected. However, as discussed in detail in the context of Fig. 5, the second harmonic is originally radiated into the heterostructure plane. As a result, second harmonic emission occurs over the entire antenna field of 100 µm x 100 µm size and even from adjacent patterns. It is well possible that second harmonic components are guided even further through the AlN layer. Consequently, the collimation of the SHG radiation using a microscope objective with a numerical aperture of 0.25 and the measurement in the far-field using the CCD in the spectrometer certainly does not collect all the SHG of the metasurface. Furthermore, the SHG might not completely couple back to the far-field in transmission geometry by the antenna array but is further guided in the AlN slab (see mode discussion below). This assumption is supported by the optical impression of the green THG light, visible to the naked eye, when looking onto the side facets of the sample, indicating good guidance inside the AlN waveguide. This leads to an underestimation of the SHG signal strength. In comparison to different realizations of metasurfaces found in literature, the introduced nonlinear metasurface reveals a power conversion efficiency above hybrid structures based on TMDCs and plasmonic antennas ($\eta_{SHG} = 10^{-13}$) [23] or plasmonic metalens ($\eta_{SHG} = 10^{-12}$) [24]. Nevertheless, metasurfaces with, e.g., optimized plasmonic three-arm trapezoidal silver antennas reach a power conversion efficiency as high as $\eta_{SHG} = 10^{-9}$ [25].

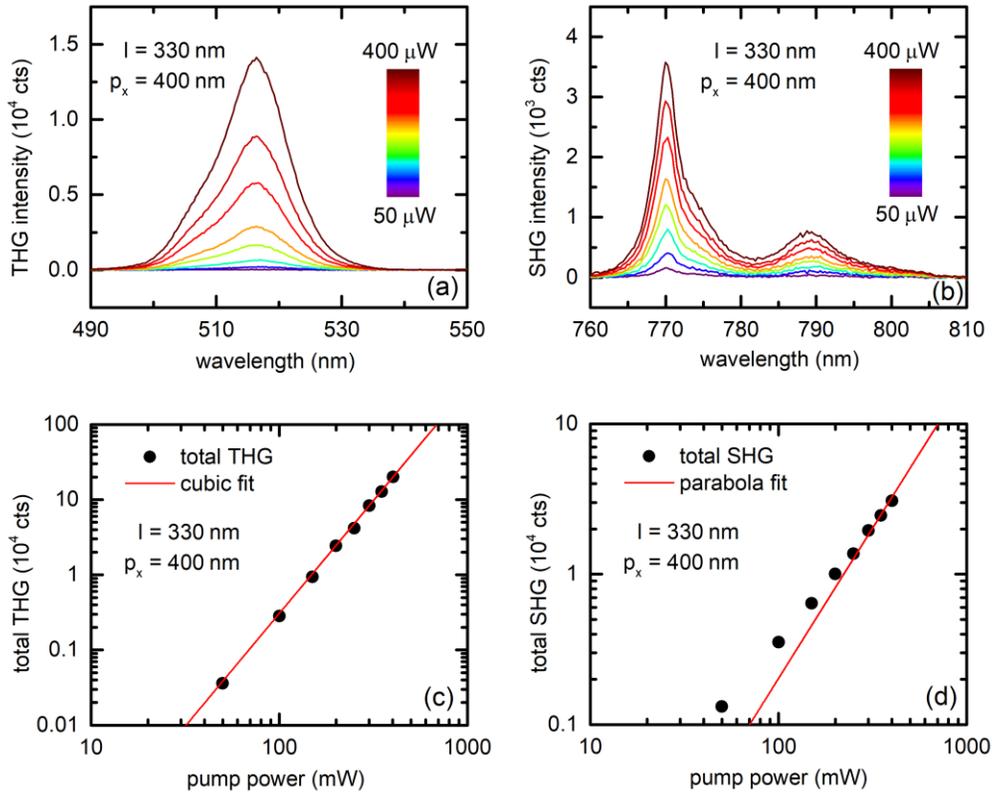

Fig. 3. (a) THG and (b) SHG spectra for various fundamental pump-powers. The antenna array on top of the GaN/AlN MQW is characterized by l = 330 nm and $p_x$ = 400 nm. (c) and (d) show the results for the spectrally integrated SHG and THG intensity ("total") together with quadratic and cubic fits, respectively.

*4.2 Polarization dependence*

We now turn towards polarization-dependent measurements at a fixed pump power. A half-wave plate in front of the sample allows for continuous rotation of the linear polarization of the fundamental beam. For longitudinal polarization (polarization parallel to the long axes of the antennas) we observe the strongest SHG, cf. Fig. 4(a). In the case of transverse polarization

(polarization axes perpendicular to the long axes of the antennas) practically no SHG takes place. We note that the data in Fig 4(a) is recorded for non-polarized detection. This observation stands in perfect agreement with the antennas' resonances originating from their geometrical shape and dimensions. Only in the case of roughly 1550 nm wavelength and longitudinal light polarization, a significant amount of the fundamental laser light is converted into z-polarized field components in the MQW, thereby allowing for SHG. In the case of transversal polarization, the antennas do not exhibit any resonance around ~1550 nm (data not shown). We note that a similar dependence on the polarization of the fundamental is also seen for THG (data not shown).

We now address the polarization state of the SHG and THG for longitudinally polarized fundamental light. To this end, a polarizer is inserted into the detection beam path. As depicted in Fig. 4 (b) the second harmonic light is cross-polarized with respect to the fundamental. We attribute this finding to the cross-polarized resonances of the antennas for fundamental and second harmonic light seen in Fig. 1. In contrast, the THG is polarized parallel to the fundamental light, as expected for rod antennas.

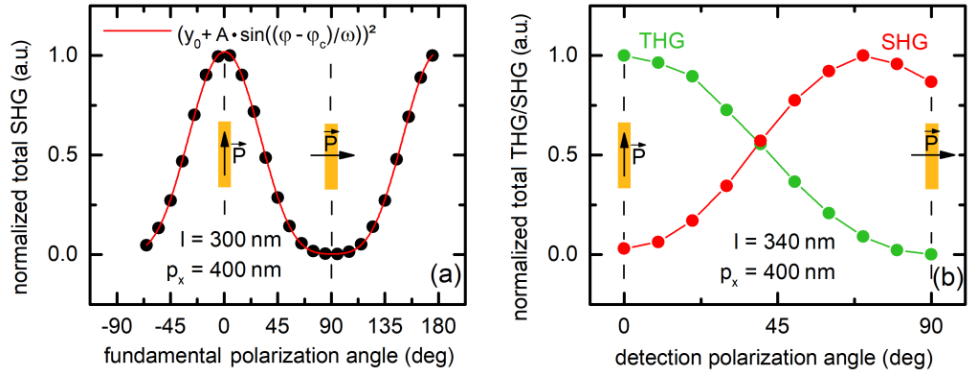

Fig. 4. (a) SHG signal as a function of the polarization angle of the fundamental light beam (0° longitudinal, 90° transversal) for non-polarized detection. The red line visualizes a fit to the data. (b) Normalized spectrally integrated SHG and THG intensity ("total") for change of the detection polarization angle (0° longitudinal, 90° transversal) for a longitudinal polarized fundamental beam.

### 4.3 Influence of the antenna periodicity

We now turn towards the most surprising aspect of our study: the observation of substantial SHG is restricted to a fairly small range of lateral antenna spacings $p_x$. Also, the visual appearance of the second harmonic emission drastically depends on $p_x$. To illustrate this aspect, an exemplary set of pictures taken for $l = 340$ nm and $p_x$ ranging from 300 nm up to 500 nm is shown in Fig. 5. Note that these pictures are rotated by 90° in comparison to the SEM picture in Fig. 1(b). For $p_x = 300$ nm and $p_x = 500$ nm, we find second harmonic emission to occur at the excitation spot. In marked contrast, the emission spot for $p_x = 350$ nm, $p_x = 400$ nm and $p_x = 450$ nm appears to be much wider. Light seems to be emitted even from the adjacent antenna patterns (pattern size: 100 μm x 100 μm) which are located perpendicular to the polarization direction of the fundamental light beam. The only explanation for this finding is that the second harmonic is radiated into the plane of the quantum wells and then converted into free-space radiation by the periodic arrangements of antennas. In particular, the plasmonic elements of the heterostructure partially convert the impinging light into z-polarized components in the MQWs. These components are aligned along the dipole moment of the ISBTs and can, therefore, be converted into the second optical harmonic when the ISBTs are strongly driven. Also, the diffraction by 90° is not completely unexpected: the second harmonic radiation has a wavelength of ~400 nm inside the heterostructure since the vacuum wavelength is ~775 nm and the refractive index is ~2. Details and the precise mode index depend on the AlN buffer layer underneath the MQWs which mostly guides the second-harmonic radiation. This guiding effect seems to be pronounced as the light even reaches adjacent antenna patterns (indicated by blue squares in the three pictures in the middle of Fig. 5), where the SHG is also scattered back into the z-direction.

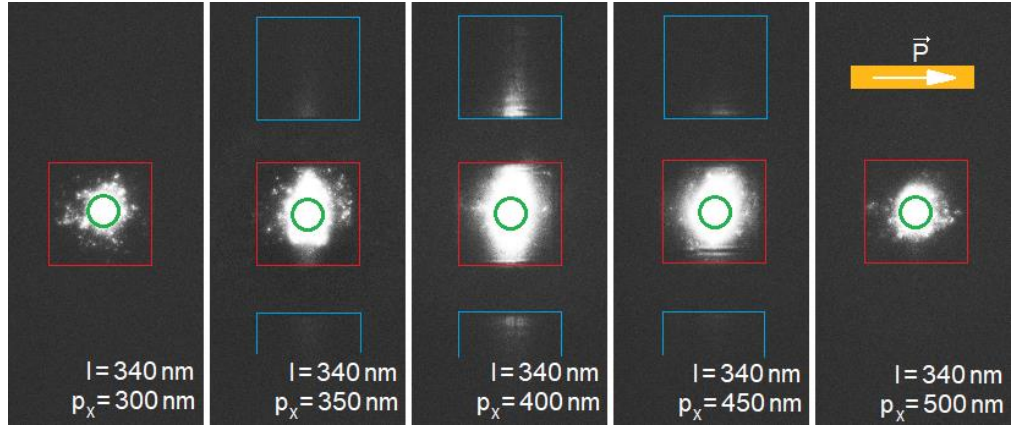

Fig. 5. Pictures taken from different illuminated antenna arrays (red boxes, array size: 100 μm x 100 μm) on top of the MQW. The longitudinally polarized pump laser (the spot on the sample is indicated by a green circle) is filtered out by a suitable shortpass thus only THG and SHG contribute to the illuminated areas. SHG can be observed for $p_x$ = 350 nm to $p_x$ = 450 with in-plane emission to adjacent antenna patterns (blue boxes). For better illustration this picture is rotated by 90 degrees with respect to the SEM picture of the sample in Fig. 1 (b).

Next, we will elaborate on the $p_x$-dependence of the emitted second and third harmonic spectra. Fig. 6(a) displays a false-color plot of the emitted third harmonic spectrum for various $p_x$. While the intensity is seen to increase with $p_x$, no significant changes to the spectrum are observed. The increase of the intensity with $p_x$ is somewhat surprising as the areal density of the antennas decreases with $p_x$ because fewer antennas contribute to the nonlinear optical emission. This observation points to a coupling of the antennas even for distances as large as $p_x$ = 500 nm. The constant spectrum points to the THG of the antenna array itself with no major impact on the optical mode structure of the semiconductor underneath.

In stark contrast, the spectrum of the emitted second harmonic in Fig. 6(b) massively depends on $p_x$ (note that the false-color plot in panel (b) uses a logarithmic scale to enhance the visibility of smaller side-peaks). For a fixed $p_x$, we typically observe emission into a few optical modes with a spectral separation of ~20 nm (c.f. also Fig. 3(b)). This finding reflects the modal structure of the ~1.6 µm thick semiconductor layer on top of the sapphire substrate. It also accounts for the observed lateral guidance of the SHG underneath the antenna array to the edges of the sample. Due to the relatively large thickness of the AlN layer, several higher-order modes can be excited by the periodic antenna arrays. Hereby the mode dispersion shifts the spectral coupling with varying $p_x$. Because the wavelength range of the SHG is restricted by the fundamental laser bandwidth, we observe an enhanced SHG signal if the generated light matches one of the TM modes of the slab (see the calculated TM mode dispersion in Fig. 6(b), indicated by red lines). The coupling of the localized plasmon resonances of the antennas along the short axis together with the stronger grating effect leads to the strong coupling for the TM modes. On the other hand, the TE modes of the slab cannot couple to the plasmon resonance and no enhanced SHG emission is observed for these modes.

To further illustrate the impact of these optical modes, Fig. 6(c) contains second harmonic spectra for a fixed period of $p_x$ = 420 nm but varying angle of incidence. Note that the sample is rotated around the longitudinal axes of the antennas. Within this small range of angles, changes of the effective lateral periodicity are practically negligible. However, the fundamental light now impinging off normal incidence provides additional in-plane momentum for the coupling of second harmonic light into the optical modes of the semiconductor heterostructure. In line with this reasoning, the modes of the emitted spectrum in Fig. 6(c) exhibit a linear shift when the angle of incidence is changed. This shift is illustrated with the dashed lines.

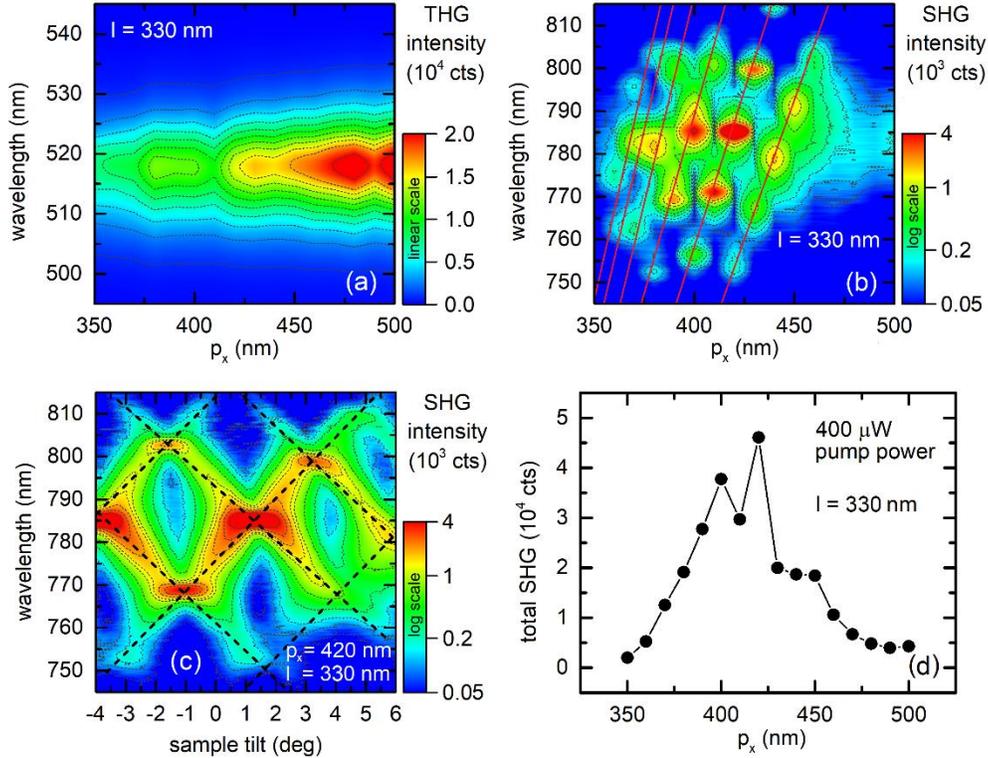

Fig. 6. (a), (b) False-color representations of the spectra of the emitted third harmonic (panel (a), linear scale) and second harmonic (panel (b), logarithmic scale) for different lateral periodicities $p_x$ of the antennas. The red lines indicate calculated higher-order TM-modes. (c) False-color representation (logarithmic scale) of the second harmonic spectra for different tilt angles of the sample with respect to the incident fundamental beam. An angle of zero degrees refers to normal incidence. The tilt occurs along the long axes of the antennas. The dotted lines visualize the observed spectral modes originating from the tilting of the sample. (d) Spectrally integrated SHG extracted from the data in panel (b).

Finally, we investigate the dependence of the spectrally integrated second harmonic on $p_x$, which is shown in Fig. 6(d). The data are obtained from the spectra shown in panel (b) and perpendicular incidence. They show a pronounced maximum of the SHG for $p_x = 420$ nm. The efficiency of the nonlinear optical process decreases fairly rapidly for both smaller and larger separations. To exclude a substantial influence of the optical mode structure, we have also recorded data sets for the spectrally integrated SHG at different small tilt angles (data not shown). It shows a practically identical dependence of the integrated SHG on $p_x$. Even the somewhat reduced SHG for $p_x = 410$ nm is present in this additional data set which might be related to imperfections of this specific antenna array. Apparently, lateral separations around $p_x = 420$ nm are best suited for SHG in the plane of the heterostructure and its conversion into free-space radiation emitted perpendicular to the surface. At this period/wavelength combination, the best mode matching to the guide modes in the AlN film for coupling perpendicular to the free-space radiation is obtained.

## 5. Conclusion

In summary, we have realized and investigated a nonlinear metasurface, taking advantage of ISBTs in ultranarrow n-doped GaN/AlN MQWs and plasmonic elements designed to be in resonance with these ISBTs. The antennas partially convert light impinging at normal incidence into field components polarized vertically to the heterostructure's plane. These fields strongly drive the ISBTs and give rise to SHG initially radiated into the plane of the quantum wells. By appropriately choosing the lateral spacing within the periodic antenna array, the second harmonic is redirected into a transmission geometry with respect to the incident driving pulse. Owing to the large conduction band offsets in the GaN/AlN material system, our metasurface exhibits SHG for fundamental pulses in the telecom window which is hard to realize in any other established semiconductor family. Remarkably, this concept allows for SHG in a nominally inversion-symmetric metasurface with a power conversion efficiency of $\eta_{SHG} = 2.5 \cdot 10^{-11}$ in the far-field. An increase in efficiency is envisioned for an optimized out-coupling of the SHG and, thereby, associated with an improved antenna geometry, making an outlook to future research.

## 6. Funding, acknowledgments, disclosures and data availability

*6.1 Funding and acknowledgments*

This work is supported by the DFG funded collaborative research center SFB TRR 142 on tailored nonlinear photonics, subprojects B02 and C05.

T. Z. acknowledges funding by the European Research Council (ERC) under the European Union's Horizon 2020 research and innovation program (grant agreement No. 724306).

V.N.J. acknowledges the partial support of MBE growth of GaN/AlN QW structures at the Ioffe Institute by the RFBR (projects #19-52-12057 DFG-a and #21-52-50004 JSPS_a).

The authors acknowledge financial support by TU Dortmund University within the funding program Open Access Publishing.

*6.2 Disclosures*

The authors declare no conflicts of interest related to this article.

*6.3 Data availability*

Data underlying the results presented in this paper are not publicly available at this time but maybe obtained from the authors upon reasonable request.

## 7. References

1. S. V. Makarov, A. N. Tsypkin, T. A. Voytova, V. A. Milichko, I. S. Mukhin, A. V. Yulin, S. E. Putilin, M. A. Baranov, A. E. Krasnok, I. A. Morozov, and P. A. Belov, "Self-adjusted all-dielectric metasurfaces for deep ultraviolet femtosecond pulse generation," Nanoscale **8**(41), 17809–17814 (2016).
2. A.V. Krasavin, P. Ginzburg, G.A. Wurtz, and A.V Zayats, "Nonlocality-driven supercontinuum white light generation in plasmonic nanostructures," Nat. Commun. **7**, 11497 (2016).
3. M. Geissbuehler, L. Bonacina, V. Shcheslavskiy, N. L. Bocchio, S. Geissbuehler, M. Leutenegger, I. Märki, J.-P. Wolf, and T. Lasser, "Nonlinear Correlation Spectroscopy (NLCS)," Nano Lett. **12**(3), 1668–1672 (2012).
4. S. Barz, G. Cronenberg, A. Zeilinger, and P. Walther, "Heralded generation of entangled photon pairs," Nature Photonics **4**, 553–556 (2010).
5. B. E. Saleh and M. C. Teich, "Fundamentals of Photonics," second ed., John Wiley & Sons, (2012).
6. R.W. Boyd, "Nonlinear Optics," fourth ed., Academic Press, (2020).

7. J. B. Khurgin, "Graphene – A rather ordinary nonlinear optical material," Appl. Phys. Lett. **104**, 161116 (2014).
8. M. Zhao, Z. Ye, R. Suzuki, Y. Ye, H. Zhu, J. Xiao, Y. Wang, Y. Iwasa, and X. Zhang, "Atomically phase-matched second-harmonic generation in a 2D crystal," Light Sci. Appl. **5** e16131 (2016).
9. J. Lee, M. Tymchenko, C. Argyropoulos, P.-Y. Chen, F. Lu, F. Demmerle, G. Boehm, M.-C. Amann, A. Alù, and M. A. Belkin, "Giant nonlinear response from plasmonic metasurfaces coupled to intersubband transitions," Nature **511**, 65-69 (2014).
10. N. Nouri and M. Zavvari, "Second-Harmonic Generation in III-Nitride Quantum Wells Enhanced by Metamaterials," Photon. Technol. Lett. **28**(20), 2199–2202 (2016).
11. L. Nevou, M. Tchernycheva, F. Julien, M. Raybaut, A. Godard, E. Rosencher, F. Guillot, and E. Monroy, "Intersubband resonant enhancement of second-harmonic generation in quantum wells," Appl. Phys. Lett. **89**, 151101 (2006).
12. M.-H. Yuan, H.Li, J.-H. Zeng, H.-H. Fan, Q.-F. Dai, S. Lan, and S.-T. Li, "Efficient blue light emission from In0.16Ga0.84N/GaN multiple quantum wells excited by 2.48-μm femtosecond laser pulses," Optics Letters **39**(12), 3555 (2014).
13. A. Krasnok, M. Tymchenko, and A. Alù, "Nonlinear metasurfaces: a paradigm shift in nonlinear optics," Materials Today, **21**(1d), 8–21 (2018).
14. G. Martin, A. Botchkarev, A. Rockett, and H. Morkoc, "Valence-band discontinuities of wurtzite GaN, AlN, and InN heterojunctions measured by x-ray photoemission spectroscopy," Appl. Phys. Lett. **68**(18), 2541-2543 (1996).
15. O. Wolf, A. A. Allerman, X. Ma, J. R. Wendt, A. Y. Song, E. A. Shaner, and I. Brener, "Enhanced optical nonlinearities in the near-infrared using III-nitride heterostructures coupled to metamaterials," Appl. Phys. Lett. **107**, 151108 (2015).
16. D. V. Nechaev, O. A. Koshelev, V. V. Ratnikov, P. N. Brunkov, A. V. Myasoedov, A. A. Sitnikova, S. V. Ivanov, V. N. Jmerik, "Effect of stoichiometric conditions and growth mode on threading dislocations filtering in AlN/c-Al$_2$O$_3$ templates grown by PA MBE," Superlattices and Microstructures **138**, 106368 (2020).
17. V. N. Jmerik, D. V. Nechaev, A. A. Toropov, E. A. Evropeitsev, V. I. Kozlovsky, V. P. Martovitsky, S. Rouvimov, and S. V. Ivanov, "High-efficiency electron-beam-pumped sub-240-nm ultraviolet emitters based on ultra-thin GaN/AlN multiple quantum wells grown by plasma-assisted molecular-beam epitaxy on c-Al$_2$O$_3$, Appl. Phys. Express **11**, 091003 (2018).
18. M. E. Levinshtein, S. L. Rumyantsev, and M.S. Shur, "Properties of Advanced Semiconductor Materials: GaN, AlN, InN, BN, SiC, SiGe," John Wiley & Sons, New York, (2001).
19. T. Jostmeier, T. Wecker, D. Reuter, D. J. As, and M. Betz, "Ultrafast carrier dynamics and resonant inter-miniband nonlinearity of a cubic GaN/AlN superlattice", Appl. Phys. Lett. **107**, 211101 (2015).
20. T. Utikal, T. Zentgraf, T. Paul, C. Rockstuhl, F. Lederer, M. Lippitz, and H. Giessen, "Towards the Origin of the Nonlinear Response in Hybrid Plasmonic Systems," Phys. Rev. Lett. **106**, 133901 (2011).
21. M. Protte, N. Weber, C. Golla, T. Zentgraf, and C. Meier, "Strong nonlinear optical response from ZnO by coupled and lattice-matched nanoantennas," J. Appl. Phys. **125**, 193104 (2019).
22. K. L. Vodopyanov, V. Chazapis, C. C. Phillips, B. Sung, and J. S. Harris Jr., "Intersubband absorption saturation study of narrow III-V multiple quantumwells in the λ=2.8-9 μm spectral range," Semicond. Sci. Technol. **12**, 708-714 (1997).
23. F. Spreyer, R. Zhao, L. Huang, and T. Zentgraf, "Second harmonic imaging of plasmonic Pancharatnam-Berry phase metasurfaces coupled to monolayers of WS$_2$," Nanophotonics **9**(2), 351-360 (2020).
24. C. Schlickriede, N. Waterman, B. Reineke, P. Georgi, G. Li, S. Zhang, and T. Zentgraf, "Imaging through Nonlinear Metalens Using Second Harmonic Generation," Adv. Mater. **2018**(30), 1703843 (2018).
25. H. Aouani, M. Navarro-Cia, M. Rahmani, T. P. H. Sidiropoulos, M. Hong, R. F. Oulton, and S. A. Maier, "Multiresonant Broadband Optical Antennas as Efficient Tunable Nanosources of Second Harmonic Light," Nano Lett. **12**, 4997-5002 (2012).